\documentclass[a4paper,twoside,10pt]{letter}
\usepackage{graphicx,saj,multicol,subeqnarray}
\usepackage{ wasysym }
\usepackage{longtable}
\usepackage{ gensymb }
\usepackage{rotating}
\usepackage{pdflscape}
\usepackage{ mathrsfs }

\newcommand{\HII}{H\,{\sc ii}}

\def\arcsec{\hbox{$^{\prime\prime}$}}

\def\p0{\phantom{0}}

\def\papertype{}

\def\udc{...}
\begin{document}
\baselineskip=3.1truemm
\columnsep=.5truecm
\newenvironment{lefteqnarray}{\arraycolsep=0pt\begin{eqnarray}}
{\end{eqnarray}\protect\aftergroup\ignorespaces}
\newenvironment{lefteqnarray*}{\arraycolsep=0pt\begin{eqnarray*}}
{\end{eqnarray*}\protect\aftergroup\ignorespaces}
\newenvironment{leftsubeqnarray}{\arraycolsep=0pt\begin{subeqnarray}}
{\end{subeqnarray}\protect\aftergroup\ignorespaces}
%


\markboth{\eightrm 20-CM RADIO-CONTINUUM STUDY OF THE SMC: PART III -  C{\HII} REGIONS}
{\eightrm G. F. WONG, et. al.}

{\ }

\publ

\type

{\ }


\title{New 20-cm Radio-continuum Study of the Small Magellanic Cloud: Part III -- Compact {\HII} Regions}


\authors{G. F. Wong, M. D. Filipovi\'c, E. J. Crawford, N.~F.~H. Tothill, }
\authors{A.~Y. De Horta, T.~J.~Galvin}

\vskip3mm


\address{University of Western Sydney, Locked Bag 1797, Penrith South DC, NSW 2751, AUSTRALIA}
\Email{m.filipovic}{uws.edu.au}


\dates{October XX, 2012}{November XX, 2012}

\summary{We present and discuss a new catalogue of 48 compact \HII\ regions in the Small Magellanic Cloud (SMC) and a newly created deep 1420~MHz ($\lambda$=20~cm) radio-continuum image of the N\,19 region located in the southwestern part of the SMC. The new images were created by merging 1420~MHz radio-continuum archival data from the Australian Telescope Compact Array. The majority of these detected radio compact \HII\ regions have rather flat spectral indices which indicates, as expected, that the dominant emission mechanism is of thermal nature. 
} 

\keywords{Magellanic Clouds -- Radio Continuum -- Catalogs}

\begin{multicols}{2}
{

\section{1. INTRODUCTION}

The Small Magellanic Cloud (SMC), with its well established distance ($\sim$60~kpc; Hilditch et al.~2005) and ideal position in the sky - towards the coldest areas near the South Celestial Pole, allows observation of radio sources to be conducted without significant interference from Galactic foreground radiation. The SMC is an ideal location to study celestial objects like compact \HII\ regions (C\HII , Mezger et al. 1967), which may be difficult to study in our own and other distant galaxies.

The SMC has been surveyed at multiple radio frequencies using archival data (Crawford et al~2011 and Wong et al.~2011a, hereafter paper~I). Deep observations of SMC young stellar objects, compact \HII\ regions, Supernova Remnants (SNRs) and Planetary Nebulae (PNe) were presented in Oliveira et al. (2012), Indebetouw et al. (2004), Filipovi\'c et al. (2005), Filipovi\'c et al. (2008) and Filipovi\'c et al. (2009), respectively. A catalogue of radio-continuum point sources (Wong et al. 2012) towards the SMC was derived from images taken from Crawford et al.~(2011).

This is the third paper in this series; Paper~I presented newly developed high sensitivity and resolution images of the SMC. The second instalment (Wong et al. 2011b; hereafter paper~II) presented a point source catalogue created from the images in paper~I. In this paper, we present newly constructed images of the N\,19 region covering the southwestern part of the SMC, at $\nu$=1.4~GHz ($\lambda$=20~cm). We also present a catalogue of compact \HII\ regions sources towards the SMC. The catalogue is derived from images at 4800~MHz ($\lambda$=6~cm) and 8600~MHz ($\lambda$=3~cm) from Crawford et al. (2011), a 2370~MHz ($\lambda$ = 13~cm) mosaic image from Filipovi\'c et al. (2002), one of our SMC 20~cm mosaic radio-continuum images (Fig.~2 in Paper~I), the N\,19 images presented in this paper and an 843~MHz ($\lambda$=36~cm) MOST image (Turtle et al.~1998). 

In \S2 we describe the data used to create the N\,19 images and identify the compact \HII\ region. In \S3 we describe our source fitting and detection methods. In \S4 we present our new maps with a brief discussion, \S5 contains our conclusions, and the appendix contains the catalogue of compact \HII\ regions.

\section{2. DATA}

\subsection{2.1 SMC Mosaic Radio-continuum Images}

The 3 and 6~cm images (Fig.~3 and Fig.~1 in Crawford et al. 2011) were created by combining data from various ATCA projects that covered the SMC (Table 1 in Crawford et al.~2011). 
The 3 and 6 cm maps have resolutions of $\sim$20\arcsec\ and $\sim$30\arcsec\, and r.m.s.\ noise of 0.8 and 0.7~mJy/beam respectively. The 13~cm radio-continuum catalogue was produced from a SMC mosaic radio survey of 20 square degrees (Filipovi\'c et al. 2002). These observations have a beam size of $\sim$40\arcsec\ and r.m.s.\ noise of 0.4~mJy/beam. The 20~cm mosaic image (Fig.~2 in Paper~I) was created by combining data from ATCA project C1288 (Mao et al.~2008)  with data obtained for a Parkes radio-continuum study of the SMC (Filipovi\'c et al. 1997). This image has a beam size of 17.8\arcsec $\times$ 12.2\arcsec\ with r.m.s.\ noise of 0.7~mJy/beam.

The 36~cm image comes from the MOST radio survey of 36 square degrees containing the SMC field (Turtle et al.~1998). These observations have a beam size of $\sim$45\arcsec\ and r.m.s.\ noise of 0.7~mJy/beam --- equal to that of the 20~cm image.

Table~1 gives the field size and central position of all images used to derive the compact \HII\ region catalogue contained in this paper.

\centerline{{\bf Table 1.}  Field size and central position of }
\centerline{ SMC images used.}
\vskip1mm
\centerline{
\begin{tabular}{ccccccccc}
\hline
\emph{Image} &\emph{RA} &\emph{ Dec}&Field Size& \\
\hline
3~cm&01:00:00 &-73:00:00& 5\degree $\times$ 5\degree & &\\
6~cm&01:00:00 &-73:00:00 & 5\degree $\times$ 5\degree& &\\
13~cm&01:00:00 &-72:50:00 &5\degree $\times$ 4\degree & &\\
20~cm&01:00:00 &-72:00:30 & 7\degree $\times$ 9\degree & &\\
36~cm&01:00:00 &-72:30:30 & 6\degree $\times$ 6\degree & &\\
\hline
\end{tabular}}

\subsection{2.2 The SMC N\,77 region}

Observations were conducted with ATCA (project C281) over two 12 hour sessions on 25th August 1993 and 10th February 1994. Two array configurations at 20 and 13~cm ($\nu$=1377/2377 MHz) were used -- 1.5B and 6B. More details about these observations can be found in Ye et al. (1995) and Boji{\v c}i{\' c} et al. (2010).
\subsection{2.3 The SMC N\,19 region}

\subsection{2.3.1 Image Creation }

In order to create high-fidelity and high-resolution radio-continuum images of the SMC N\,19 region, we searched the Australia Telescope Online Archive\footnote{http://atoa.atnf.csiro.au} (ATOA), identifying three complementary ATCA observations that covered N\,19: projects C468, C882 and C1607. The source 1934-638 was used as the primary calibrator and 0252-712 as the secondary calibrator for all ATCA SMC observations. A brief summary of the three ATCA projects is shown in Table~2.

The software packages \textsc{miriad} (Sault and Killeen 2010) and \textsc{karma} (Gooch 2006) were used for the data reduction and analysis. Initial high-resolution images were produced from the full dataset using the \textsc{miriad} multi-frequency synthesis (Sault and Wieringa 1994) with natural weighting. The deconvolution process used \textsc{miriad} tasks \textsc{mossdi}, an SDI variant of the clean algorithm designed for mosaics (Steer et al. 1984). 

Figs.~1-3 show maps from individual ATCA projects (Table~2), Figs.~4 and 5 show maps derived from combining multiple observations.

\centerline{{\bf Table 2.} ATOA data used to image N\,19}
\vskip1mm
\centerline{
\begin{tabular}{llcccc}
\hline
\emph{ATCA}&\emph{Date}& \emph{Array}\\
\emph{Project}&\emph{Observed}& \\
\hline
C468 & 1997 Aug 06--07& 375\\
& 1995 Oct 26--27 & 1.5D\\
& 1997 Nov 22 & 6C\\
C882 & 2000 Jun 20--17 & 6B\\
C1607& 2006 Dec 02--12  & 6B \\
& 2006 Dec 12--18 & 750A \\
\hline
\end{tabular}}
\vskip.2cm

\subsection{2.3.2 Images}

Comparing individual maps of N\,19 (Figs.~1--3), we can see the effects of different array configurations. Fig.~1 is created from project C468, containing a combination of extended and point source emission, as a result of three different array configurations. Fig.~2 (project C882) only contains a long-baseline observation (array configuration 6B), so the map is dominated by point sources. Fig.~3 has extended and point source emission, derived from short and long-baseline array configurations 750A and 6B respectively. Table~3 lists the details of the individual maps.


%
%
%

\centerline{{\bf Table 3.} Details of N\,19 mosaics at 20-cm.}
\vskip1mm
\centerline{
\begin{tabular}{lccccc}
\hline
\emph{ATCA}&\emph{Beam Size}&\emph{PA} &\emph{r.m.s.\ noise}\\
\emph{Project}& (arcsec) &   & (mJy/beam)\\
\hline
C468  & 5.3$\times$5.1 &2.7\degree & 0.1\\
C882 & 6.6$\times$6.2 &-1.3\degree   & 0.1\\
C1607 & 6.9$\times$5.5 &-2.8\degree  & 0.1\\
\hline
\end{tabular}}
\vskip.5cm

Figs.~4 and 5 show images created from a combination of observations (Table~2): Fig.~4 was created from ATCA projects C468 and C1607 while Fig.~5 contains observations C468, C882 and C1607. The images contain a combination of point sources and extended emission.

}
\end{multicols}

\centerline{\includegraphics[width=0.5\textwidth,angle=-90]{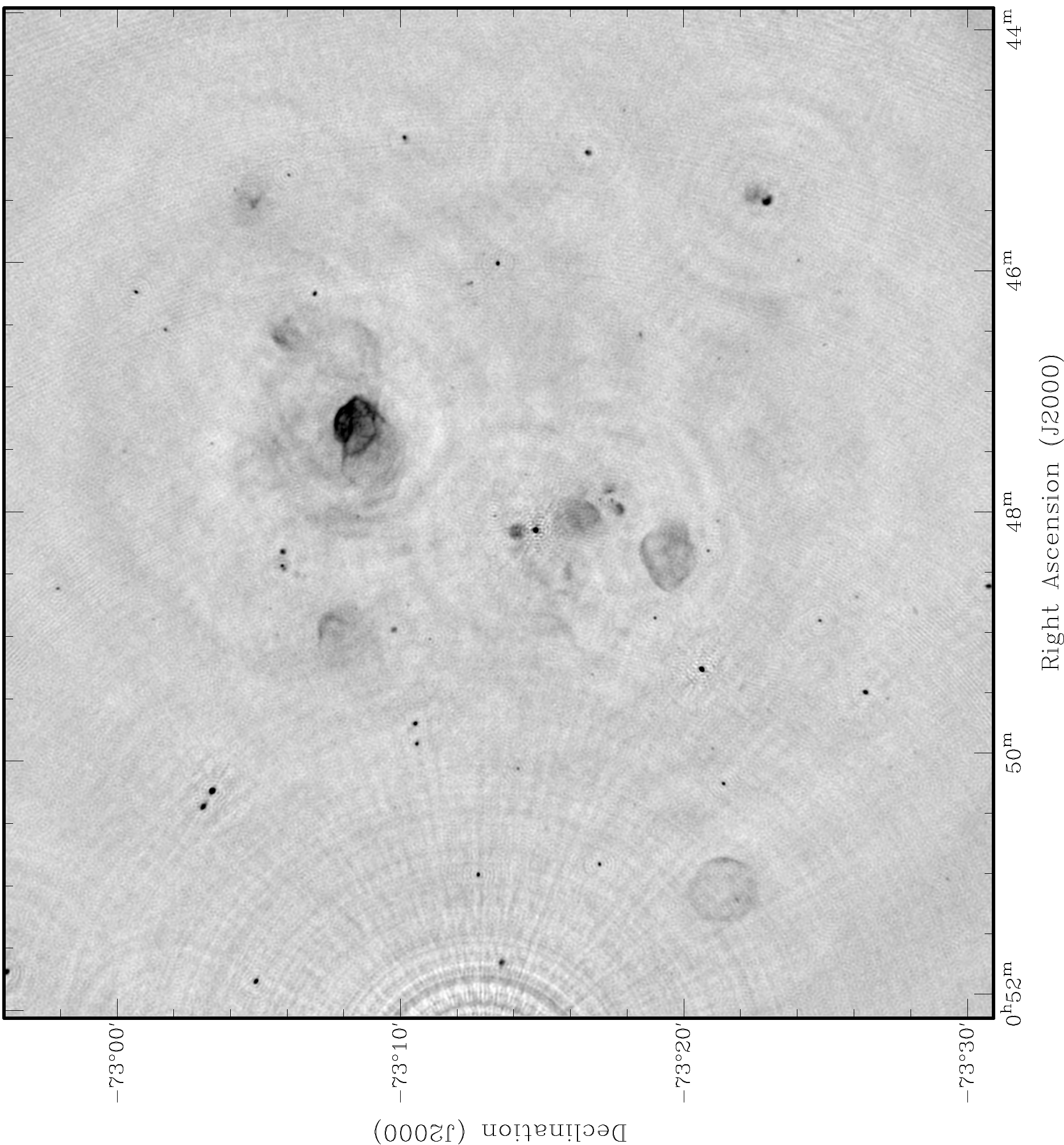}}
\figurecaption{1.}{ATCA project C468 radio-continuum total intensity image of N\,19. The synthesised beam is  5.3\arcsec$\times$5.1\arcsec\ and the r.m.s.\ noise is $\sim$0.1 mJy/beam.}

\centerline{\includegraphics[width=0.5\textwidth,angle=-90]{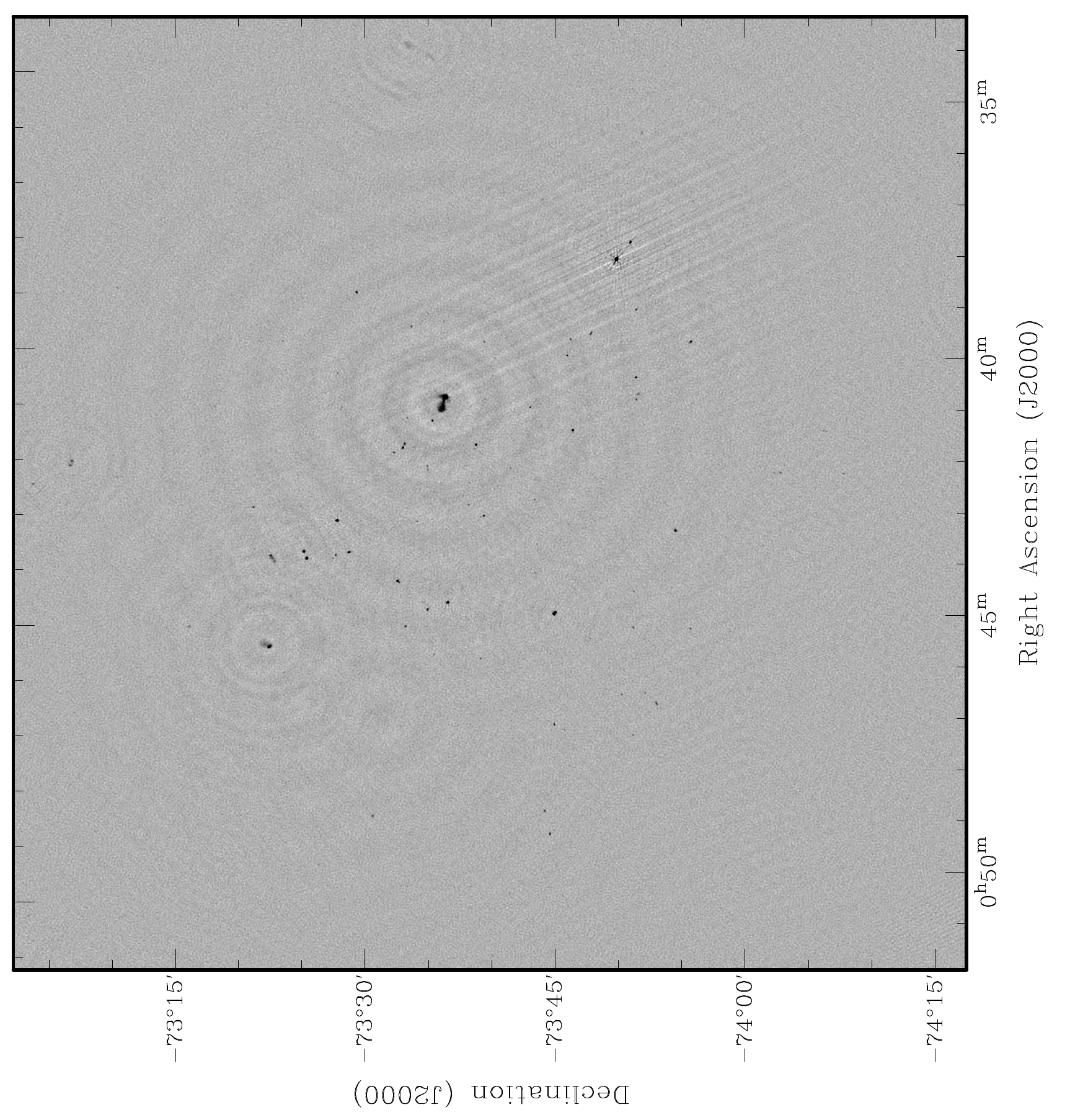}}
\figurecaption{2.}{ATCA project C882 radio-continuum total intensity image of N\,19. The synthesised beam is 6.6\arcsec$\times$6.2\arcsec\ and the r.m.s.\ noise is $\sim$0.1 mJy/beam.}

\centerline{\includegraphics[width=0.55\textwidth,angle=-90]{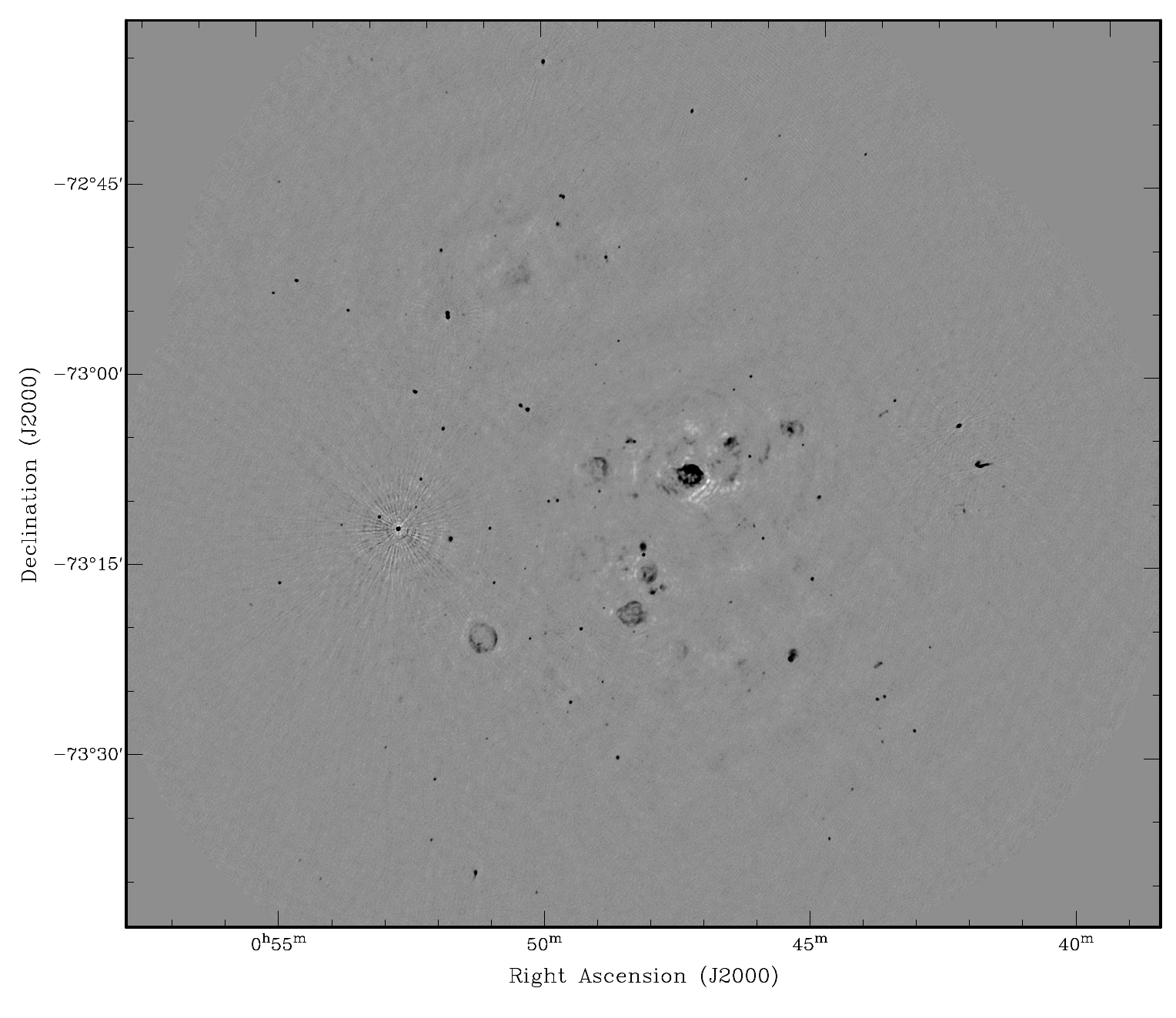}}
\figurecaption{3.}{ATCA project C1607 radio-continuum total intensity image of N\,19. The synthesised beam is 6.9\arcsec$\times$5.5\arcsec\ and the r.m.s.\ noise is $\sim$0.1 mJy/beam.}

\centerline{\includegraphics[width=0.6\textwidth,angle=-90]{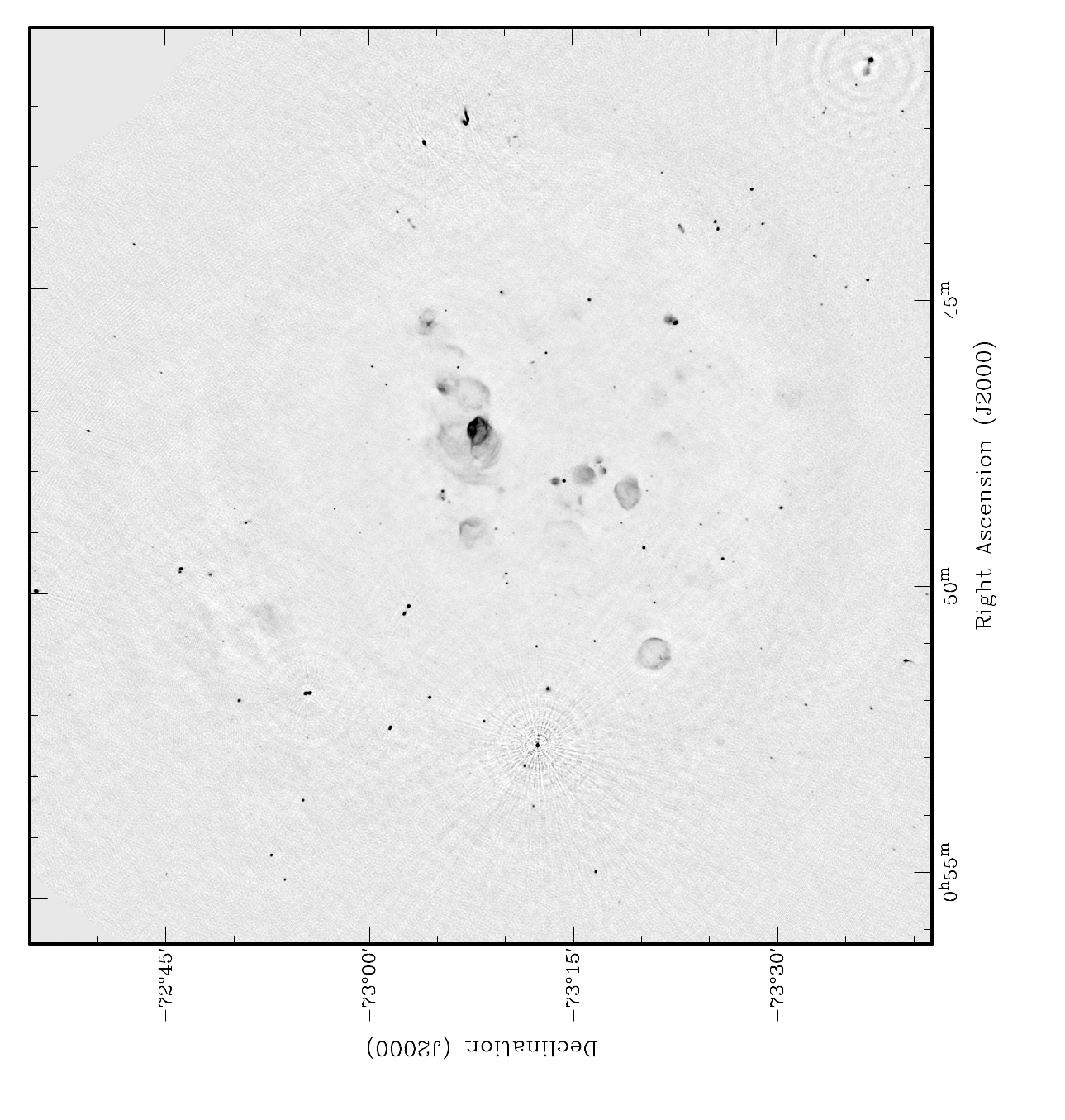}}
\figurecaption{4.}{Combined ATCA projects C468 and C1607 radio-continuum total intensity image of N\,19. The synthesised beam is 5.3\arcsec$\times$5.0\arcsec\ and the r.m.s.\ noise is $\sim$0.1 mJy/beam.}

\centerline{\includegraphics[width=0.9\textwidth,angle=-90]{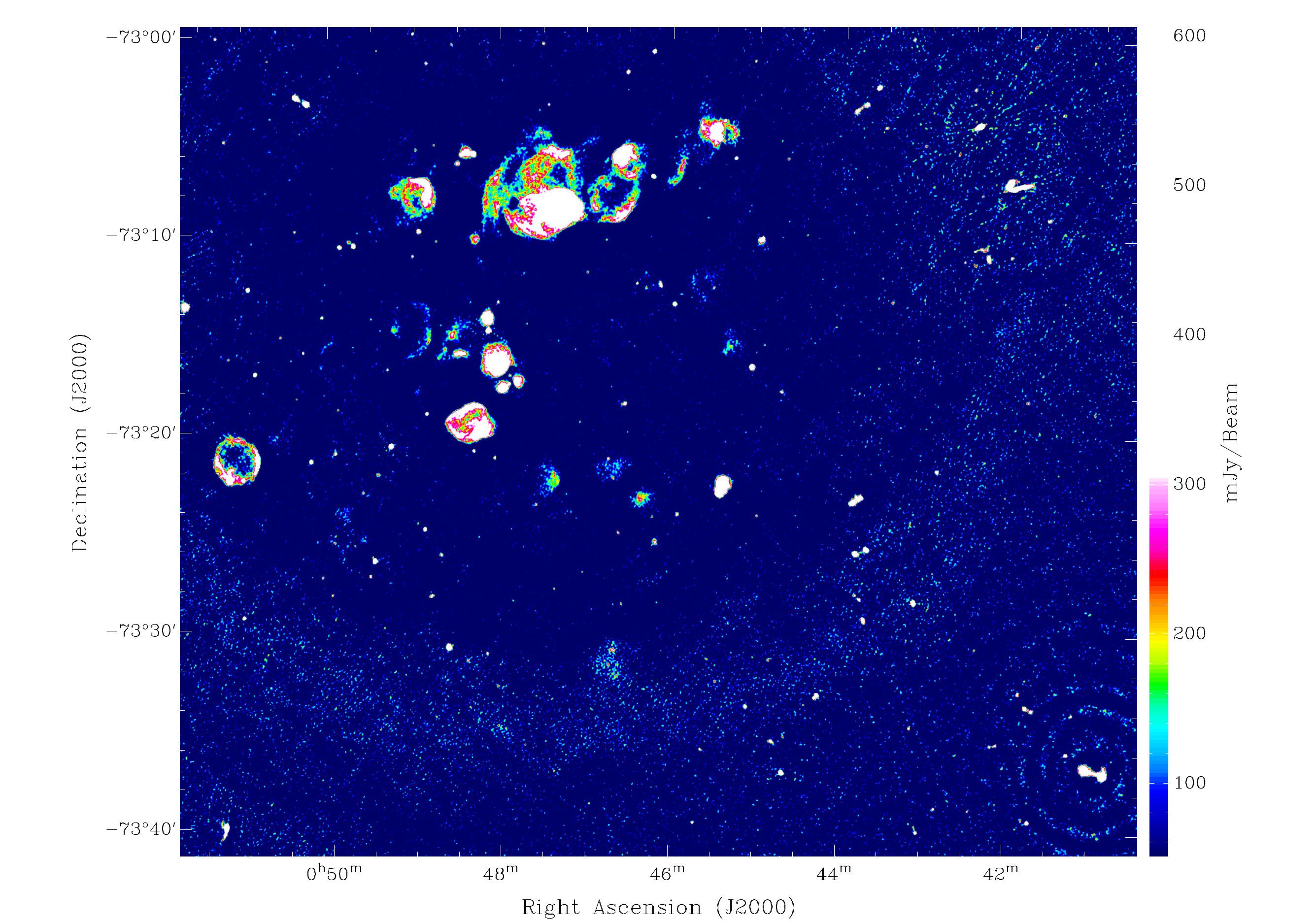}}
\figurecaption{5.}{20~cm total intensity continuum image of N\,19, derived from ATCA projects C468, C882 and C1607. The synthesised beam is 5.3\arcsec$\times$5.0\arcsec\ and the r.m.s.\ noise is $\sim$0.1 mJy/beam.}

\begin{multicols}{2}
{

\section{3. THE SMC COMPACT \HII\ REGIONS: SOURCE FITTING AND DETECTION}


Compact \HII\ regions\ were identified by comparing our list of radio-continuum point sources (Paper~II and Wong et al. 2012) to an H$\alpha$ observation from the Magellanic Cloud Emission Line Survey (MCELS; Smith et al. 1999). Sources were examined visually to confirm that matched H$\alpha$ sources were point or point-like in H$\alpha$ and within $\sim$5\arcsec of the radio point source detection.

The list of candidate compact \HII\ regions was narrowed down by cross-checking with catalogues of known objects like supernova remnants (Filipovi\'c et al. 2008) and planetary nebulae (Filipovi\'c et al. 2009, Crawford et al. 2012, Boji{\v c}i{\' c} et al. 2010).
%

Table~A1 gives the names, positions (J2000, derived from the 1420~MHz image of the whole SMC) and the integrated flux values at various frequencies for the compact \HII\ regions. The integrated fluxes in columns 5--12 are: flux density values taken from the MOST image; 1420~MHz flux density measurements from SMC (high resolution Fig.~2, Paper~1) and new N\,19 (Fig.~5) mosaic images; flux measurements retrieved from the N\,77 20 \& 13~cm images (Ye et al. 1995); flux values from an SMC 2370~MHz mosaic image (Filipovi\'c et al. 2002); 4800 and 8640~MHz flux values from Crawford et al.~(2011; Figs.~3 and 1).

\section{4. RESULTS AND DISCUSSION}

Figs.~1--3 show the individual intensity mosaic maps of the N\,19 region derived from projects C468, C882 and C1607 respectively, while Figs.~4 and 5 are images created by combining different observations. All these images can be downloaded from: spacescience.uws.edu.au/mc/smc/N19/.  A sample of 48 compact \HII\ regions was selected; table~A1 lists the compact \HII\ regions, with integrated flux values at various frequencies. These flux values are derived from gaussians fitted to the images described in \S2. 


30 of the 48 catalogue sources were detected at more than one wavelength. Spectral indices $\alpha$ ($S_\nu \propto \nu^\alpha$), with errors, were estimated for all of these sources, fitted to all available flux measurements. Integrated fluxes at 1~GHz were also derived from these fits. Integrating the fitted spectra from 10~MHz to 100~GHz yielded fluxes (in $10^{-26}$\,W\,m$^{-2}$) which were then converted into radio luminosities using the known distance to the SMC. Table~A2 lists for all compact \HII\ regions: The estimated flux density at 1~GHz; spectral index with errors; and the radio luminosity (i.e. the luminosity of the compact \HII\ region over the radio spectrum) in units of $10^{26}$\,W and in solar luminosities.


Figs.~6--8 contain spectral index distributions of the compact \HII\ regions in the N\,19 region, in the SMC (without N\,19), and the combination of the two. We expect compact \HII\ regions to have flat radio spectra ($\alpha\approx -0.1$), turning into $\alpha\approx 2$ below a cutoff frequency (which depends on the individual region); the spectral index distributions are indeed peaked around small values of $\alpha$. While the spectral index distribution of SMC sources has a broad peak with several sources having quite large $\alpha$, the distribution of N\,19 sources is much more sharply peaked near $\alpha=0$. The maps of N\,19 have better resolution and sensitivity than the maps of the rest of the SMC, so the N\,19 sources might be expected to have more accurate fluxes and hence more accurate spectral indices. 
Overall, $\sim$60\% of the sources have spectral index within the expected range of $-0.3<\alpha <-0.3$. Higher spectral indices may be explained by the cutoff frequency lying within the frequency range of our observations (see notes on individual sources below). There are also a few sources with significantly negative spectral indices, which is not consistent with thermal emission --- these may be contaminating non-\HII\ region sources.


\centerline{\includegraphics[width=0.5\textwidth]{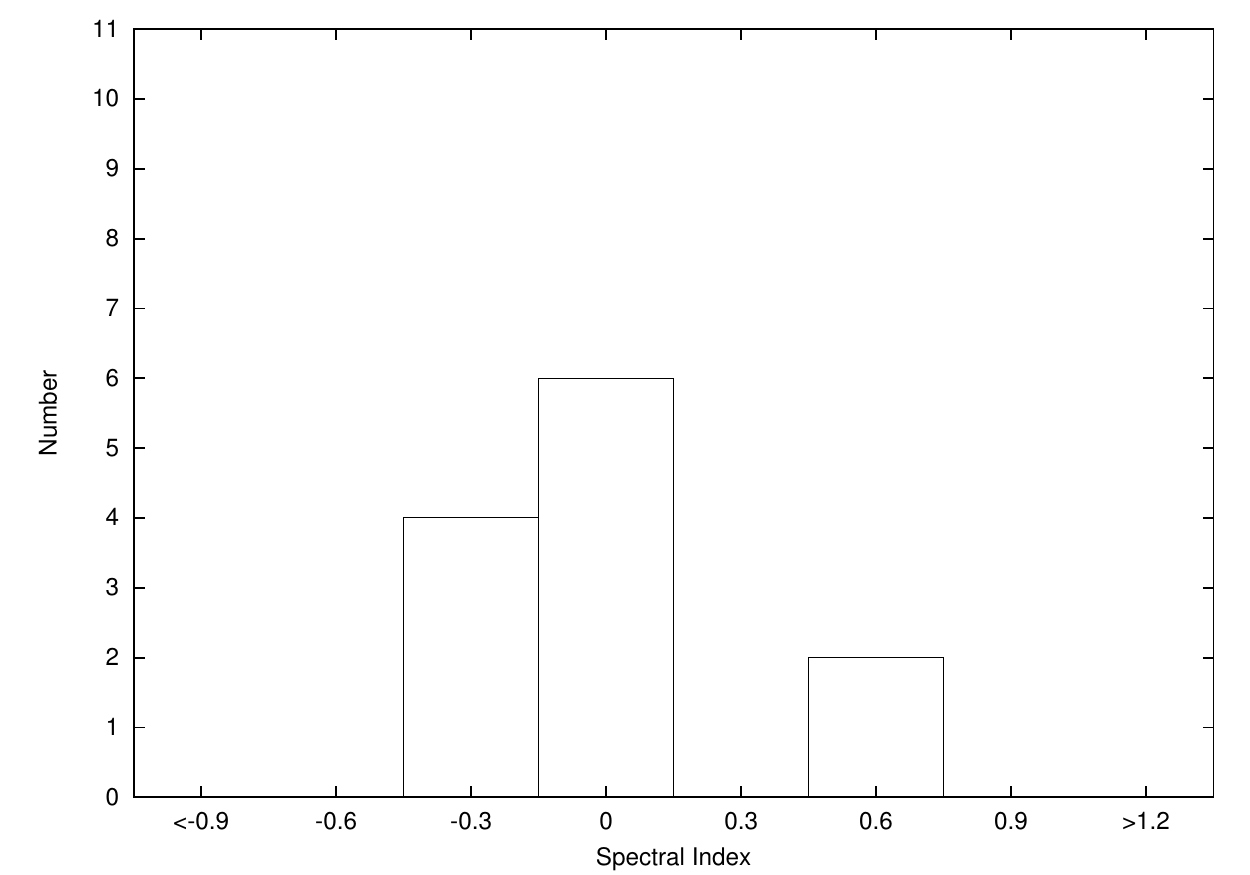}}
\figurecaption{6.}{Spectral Index distribution of compact \HII\ regions in N\,19.}

\centerline{\includegraphics[width=0.5\textwidth]{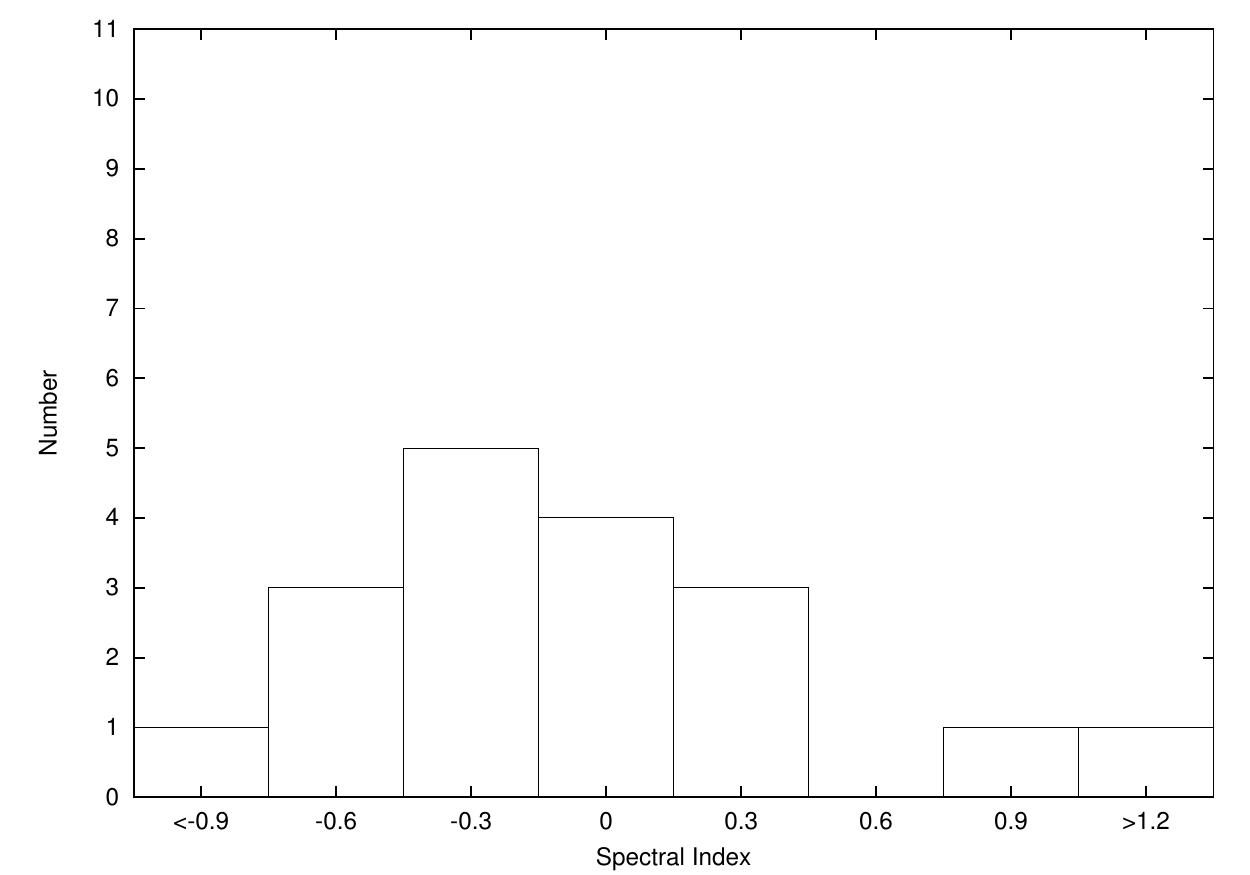}}
\figurecaption{7.}{Spectral Index distribution of SMC compact \HII\ regions (excluding N\,19 sources).}

\centerline{\includegraphics[width=0.5\textwidth]{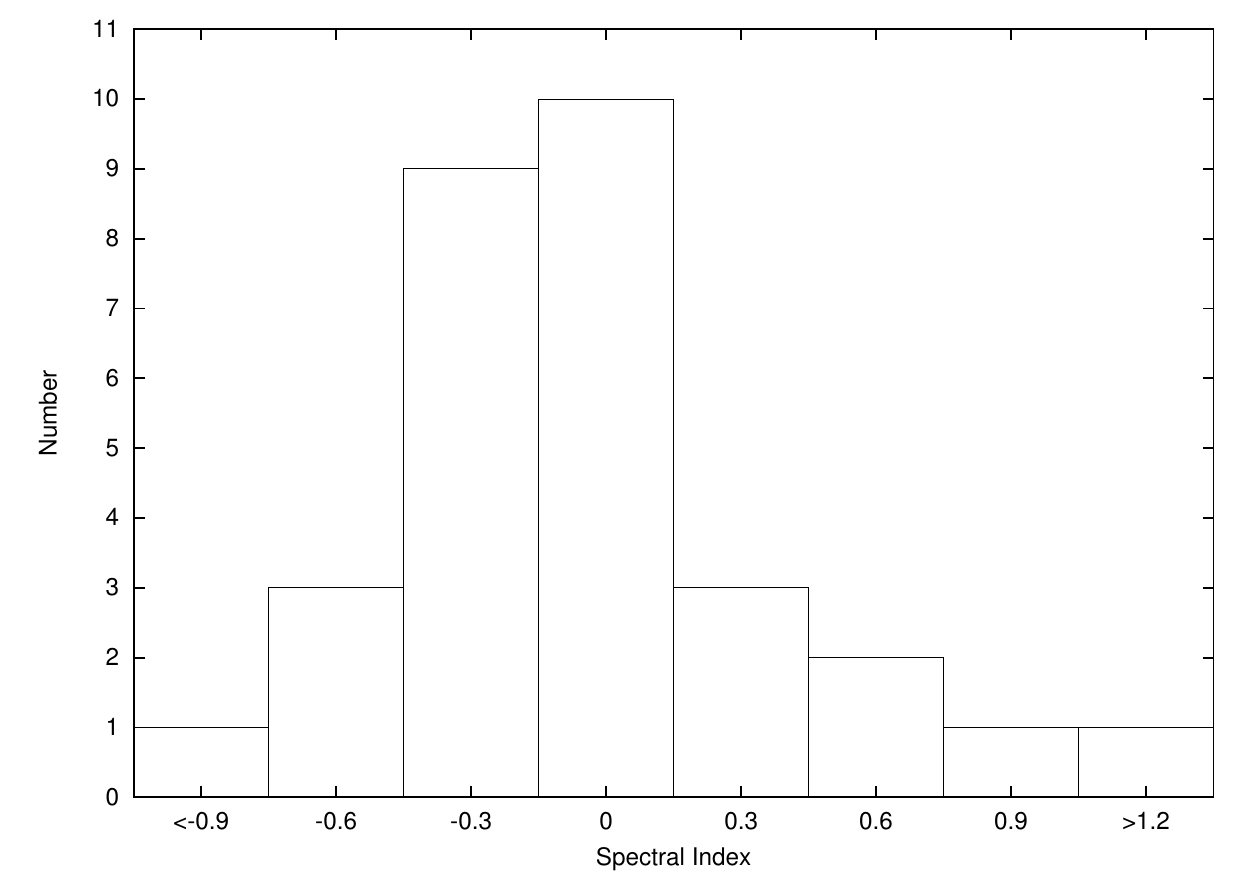}}
\figurecaption{8.}{Spectral Index distribution of compact \HII\ regions in SMC (including N\,19 sources).}

The radio luminosity distribution  (Fig.~9) peaks around $\sim$0.3\,$\mathscr{L}_\odot$: It falls off towards higher lumninosities as there are fewer sufficiently bright stars to power compact \HII\ regions at such high luminosities; it falls off towards lower luminosities due to the lack of completeness of our catalogue. Our catalogue is likely to be quite complete down to a limiting radio luminosity of $\sim$0.3\,$\mathscr{L}_\odot$.

\subsection{4.1 Notes on individual sources}

\subsection{4.1.1 J005914-721103 (Source 33)}

This compact \HII\ region candidate is the extreme positive outlier in Figs.~7\,\&\,8, with a spectral index of $+2.0\pm 0.4$; it is only detected in the 13\,cm and 20\,cm N\,77 images, so the spectral index is defined purely by these fluxes.  It is possible that this is a compact \HII\ region with a very high-frequency spectral break, so that it has a spectral index of 2 even in the higher-frequency maps. However, in the 13\,cm image, the source is located at the edge of the primary beam, so that the 13\,cm flux is rather unreliable, and so this spectral index may not be accurate. It is also possible that this source could be a variable background galaxy that flared during the N\,77 observations, and is otherwise undetected. The very high estimated luminosity is largely due to the extreme spectral index, and is unlikely to be accurate --- if the source is indeed a compact \HII\ region with a very high cutoff frequency, the high luminosity is due to extrapolation of the low-frequency spectral index to the whole radio region.

%
\subsection{4.1.2 J010132-715042 (Source 36)}

This is the other source with a highly positive spectral index ($+1.0\pm 0.4$), and could only be detected in the 20\,cm and 36\,cm images. It is likely that this is a compact \HII\ region whose spectral break lies between 36\,cm and 20\,cm. Fitting a single power-law to these data then gives an average spectral index of about 1. This will also lead to a large overestimate in the luminosity.

\subsection{4.1.3 J012408-730904 (Source 48)}

This compact \HII\ region candidate has a very high radio luminosity of 23.5 $\mathscr{L}_\odot$; this estimate comes from a combination of a positive estimated spectral index (which implies rising flux through the GHz spectrum, and hence a large luminosity) and very high fluxes in all images. The source is detected at 36\,cm, and the higher-frequency data yield a spectral index of $\sim-0.2\pm0.1$. This is quite consistent with the source being a compact \HII\ region with a cutoff frequency between 843\,MHz and 1.4\,GHz. Although the luminosity may be overestimated, this is clearly a very luminous radio source. If it is a compact \HII\ region, it is likely to be powered by a cluster of young high-mass stars, which can supply such a large luminosity.


\centerline{\includegraphics[width=0.5\textwidth]{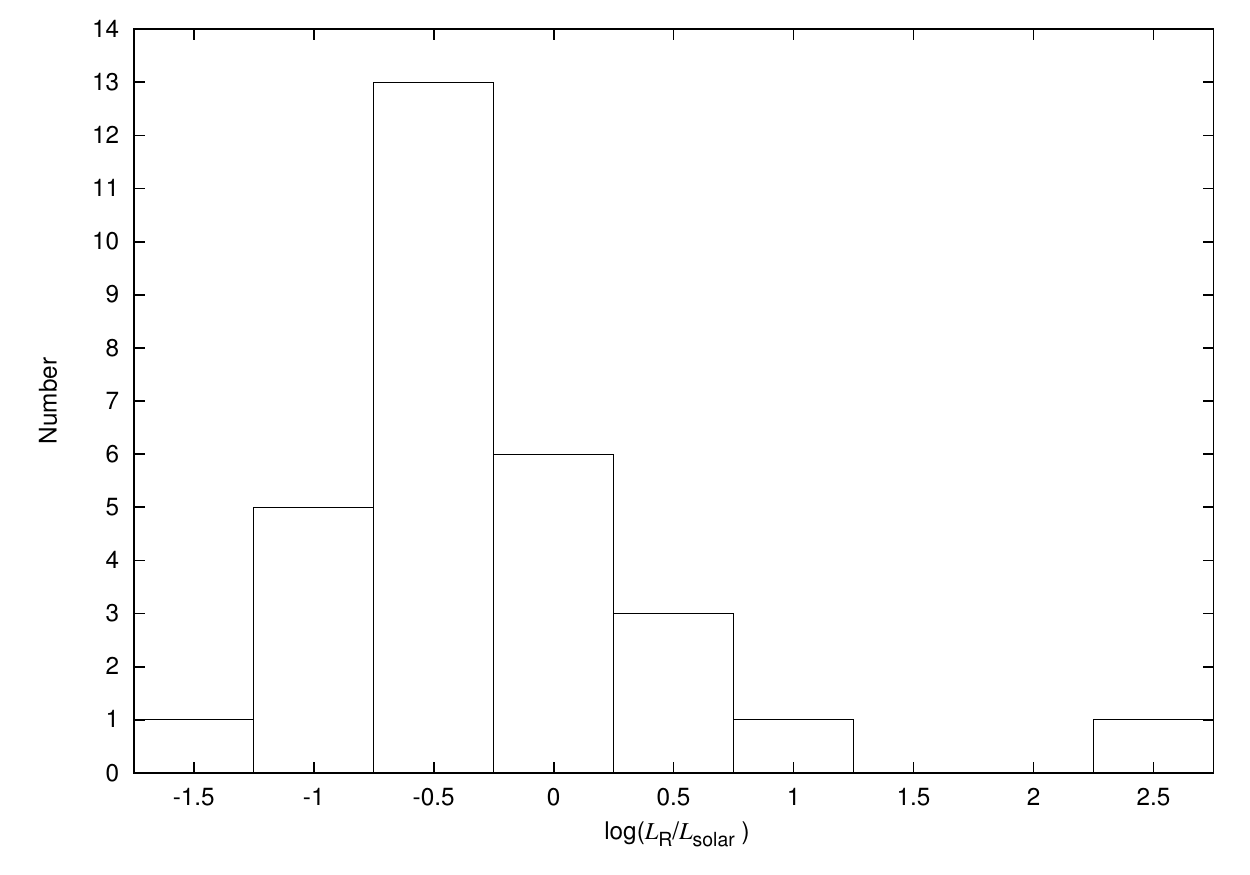}}
\figurecaption{9.}{Radio luminosity distribution of SMC compact \HII\ regions from all maps.}

}
\end{multicols}

\centerline{\includegraphics[width=0.75\textwidth,angle=-90]{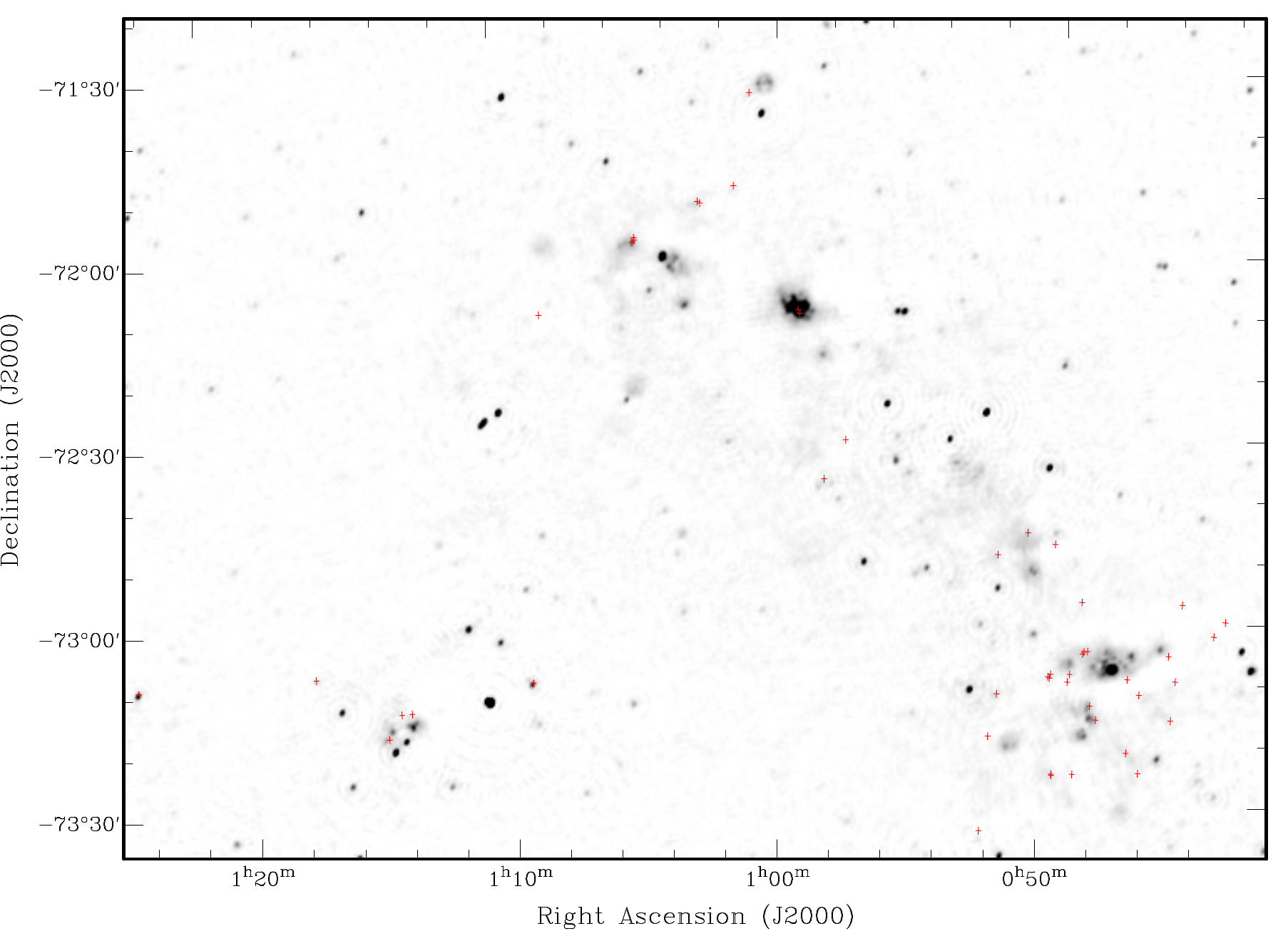}}
\figurecaption{10.}{Location of compact \HII\ regions overlaid on a continuum map of the SMC (Fig~6, Paper~I).}

\begin{multicols}{2}
{

\section{5. CONCLUSION}

In this paper we present a new catalogue of 48 candidate compact \HII\ regions within the SMC. This catalogue is derived from previously-presented datasets, and from a new set of high-sensitivity and high-resolution radio-continuum images of the N\,19 region at 1420~MHz ($\lambda$=20~cm), created from archival ATCA data. We have collected flux measurements at as many wavelengths as possible for all sources, and used these fluxes to fit a power-law radio spectrum, which in turn yields estimates of the flux at 1\,GHz and the spectral index. The distribution of spectral indices is consistent with a population of sources dominated by compact \HII\ regions, i.e.\ peaked around $\alpha\approx 0$, with better data yielding sharper peaks, and a few sources with significantly positive spectral indices --- consistent with compact \HII\ regions whose cutoff frequencies lie within our frequency range. We have also estimated the radio luminosity of the sources: The distribution of radio luminosities is strongly peaked around a presumed threshold luminosity, but includes some very luminous regions, such as J012408-730904, whose very high radio luminosity suggests that it is powered by a cluster of young high-mass stars.





\acknowledgements{The Australia Telescope Compact Array is part of the Australia Telescope National Facility which is funded by the Commonwealth of Australia for operation as a National Facility managed by CSIRO. This paper includes archived data obtained through the Australia Telescope Online Archive (http://atoa.atnf.csiro.au). We used the {\sc karma} and {\sc miriad} software packages developed by ATNF. We thank the anonymous referee and Richard Sturm for their valuable comments, which have led to an improved paper.}


\references

Boji{\v c}i{\' c}, I. S., Filipovi{\'c}, M. D., Crawford, E. J.: 2010 \journal{Serb. Astron. J.}, \vol{181}, 63.


Dickel, J.R., Gruendl, R.A.; McIntyre, V.J., Shaun W.A.: 2010, \journal{Astron. J.}, \vol{140}, 1511.

Crawford, E. J., Filipovi{\'c}, M. D., de Horta, A. Y., Wong, G. F., Tothill, N. F. H., Dra\v skovi{\'c},, D., Collier, J. D., Galvin, T. J.:2011 \journal{Serb. Astron. J.}, \vol{183}, 95.

Crawford, E.J., Filipovi{\'c}, M.D., Boji\v ci{\'c}, I.S., Cohen, M., Payne, J.L., De Horta, A.Y., Reid, W.:2012, \journal{IAU Symposium Vol. 283 of IAU Symposium}

Filipovi{\'c}, M.D., Jones, P.A., White, G.L, Haynes, R.F, Klein, U., Wielebinski, R.: 1997, \journal{Astron. Astrophys. Suppl. Series}, \vol{121}, 321.

Filipovi{\'c}, M.D., Haynes, R.F., White, G.L., Jones, P.A.: 1998, \journal{Astron. Astrophys. Suppl. Series}, \vol{130}, 421.

Filipovi{\'c}, M.D., Bohlsen, T., Reid, W, Staveley-Smith, L., Jones, P.A, Nohejl, K., Goldstein, G.: 2002, \journal{Mon. Not. R. Astron. Soc.}, \vol{335}, 1085.

Filipovi{\'c}, M.D., Payne, J.L., Reid, W., Danforth, C.W., Staveley-Smith, L., Jones, P.A., White, G.L.: 2005, \journal{Mon. Not. R. Astron. Soc.}, \vol{364}, 217.

Filipovi{\'c}, M.D., Haberl, F., Winkler, P.F., Pietsch, W., Payne, J.L., Crawford, E.J., de Horta, A.Y., Stootman, F.H., Reaser, B.E.: 2008, \journal{Astron. Astrophys.}, \vol{485}, 63.

Filipovi{\'c}, M.D., Cohen, M., Reid, W.A., Payne, J.L., Parker, Q.A., Crawford, E.J., Boji\v ci\'c, I.S., de Horta, A.Y., Hughes, A., Dickel, J., Stootman, F.: 2009, \journal{Mon. Not. R. Astron. Soc.}, \vol{399}, 769.



Gooch, R.: 2008, \textsc{Karma} Users Manual, ATNF, Sydney.

Hilditch, R.W., Howarth, I.D., Harries, T.J.: 2005, \journal{Mon. Not. R. Astron. Soc.}, \vol{357}, 304.

Indebetouw, R., Johnson, K. E., Conti, P.: 2004, \journal{Astron. J.}, \vol{128}, 2206.

Mao, S.A., Gaensler, B.M., Stanimirovi{\'c}, S., Haverkorn, M., McClure-Griffiths, N.M., Staveley-Smith, L., Dickey, J.M.: 2008, \journal{Astrophys. J.}, \vol{688}, 1029.


Mezger, P. G., Altenhoff, W., Schraml, J., Burke, B. F., Reifenstein, E. C., III, Wilson, T. L.:1967 \journal{Astron. J.}, \vol{150}, L157.

Oliveira, J.~M., Th. van Loon, J., Sloan, G.C., Sewilo, M., Kraemer, K.E.,Wood, P.R., Indebetouw, R., Filipovi{\'c}, M.D., Crawford, E. J., Wong, G. F., Hora, J.L., Meixner M., Robitaille, T.P., Shiao, B. Simon, J.D.:2012, \journal{Mon. Not. R. Astron. Soc.}, (in press).

Sault, R., Killeen, N.: 2010, Miriad Users Guide, ATNF

Sault, R.J., Wieringa, M.H.: 1994, \journal{Astron. Astrophys. Suppl. Series}, \vol{108}, 585.

Smith,R.C. and MCELS Team.:1999, \journal{IAU Symposium Vol. 190 of IAU Symposium}

Steer, D.G., Dewdney, P.E., Ito, M.R.: 1984, \journal{Astron. Astrophys.}, \vol{137}, 159.

Turtle, A.J., Ye, T., Amy,~S.W., Nicholls,~J.: 1998, \journal{Publ. Astron. Soc. Aust}, \vol{15}, 280.

Wong, G.F., Filipovi{\'c}, M.D., Crawford, E.J., de Horta, A.Y., Galvin, T., Dra\v skovi{\'c}, D., Payne, J.L.: 2011a, \journal{Serb. Astron. J.}, \vol{182}, 43.

Wong, G. F., Filipovi{\'c}, M. D., Crawford, E. J., Tothill, N. F. H., de Horta, A. Y.,  Dra\v skovi{\'c}, D., Galvin, T. J., Collier, J. D., Payne, J. L.: 2011b, \journal{Serb. Astron. J.}, \vol{183}, 103.

Wong, G. F., Crawford, E. J., Filipovi{\'c}, M. D., De Horta, A. Y., Tothill, N. F. H., Collier, J. D.,  Dra\v skovi{\'c}, D., Galvin, T. J., Payne, J. L.: 2012, \journal{Serb. Astron. J.}, \vol{184}, 93.

Ye, T. S., Amy, S. W., Wang, Q. D., Ball, L., Dickel, J.: 1995, \journal{Mon. Not. R. Astron. Soc.}, \vol{275}, 1218.

\endreferences

}
\end{multicols}

\begin{landscape}
\section{APPENDIX}

\centerline{{\bf Table A1.}Compact \HII\ regions and their Fluxes.}
\vskip1mm
\setlongtables
\small
\begin{longtable}{cccccccccccc}
\hline
(1)& (2)         & (3)        & (4)         & (5)           & (6)        & (7)        & (8)        & (9)        & (10)  & (11) & (12)\\
No & Name    & RA         & Dec         & S$_{843}$     & S$_{1420} (SMC)$ &S$_{1420} (N\,19)$ &S$_{1377} (N77)$ & S$_{2377} (N77)$ & S$_{2370}(SMC) $ & S$_{4800}$ & S$_{8640}$\\
   &             & (J2000)    & (J2000)     & (mJy)          & (mJy)       &  (mJy)      &  (mJy)      &(mJy)   &(mJy)      &  (mJy)  &  (mJy)\\
\hline
1 & J004312-725958  & 00:43:12.6 & -72:59:58 & --- & --- & 0.55 & --- & --- & --- & --- & --- \\
2 & J004336-730227  & 00:43:36.4 & -73:02:27 & 5.9 & 5.0 & 2.44 & --- & --- & 5.2 & 5.8 & 4.8 \\
3 & J004451-725734  & 00:44:51.6 & -72:57:34 & --- & 2.3 & 0.20 & --- & --- & --- & --- & --- \\
4 & J004457-731012  & 00:44:57.1 & -73:10:12 & 5.7 & 3.0 & 2.66 & --- & --- & 4.6 & 5.2 & 3.5 \\\smallskip
5 & J004502-731639  & 00:45:02.6 & -73:16:39 & 6.3 & 6.4 & 2.61 & --- & --- & 5.4 & 4.4 & 4.5 \\
6 & J004515-730607  & 00:45:15.7 & -73:06:07 & --- & --- & 1.54 & --- & --- & --- & --- & --- \\
7 & J004610-732534  & 00:46:10.7 & -73:25:34 & --- & --- & 0.40 & --- & --- & --- & --- & --- \\
8 & J004617-731243  & 00:46:17.7 & -73:12:43 & --- & --- & 0.23 & --- & --- & --- & --- & --- \\
9 & J004640-732221  & 00:46:40.3 & -73:22:21 & --- & --- & 0.31 & --- & --- & --- & --- & --- \\\smallskip
10 & J004645-731017  & 00:46:45.9 & -73:10:17 & --- & --- & 0.19 & --- & --- & --- & --- & --- \\
11 & J004753-731709  & 00:47:53.5 & -73:17:09 & --- & 6.7 & 0.38 & --- & --- & --- & --- & --- \\
12 & J004808-731454  & 00:48:08.6 & -73:14:54 & 19.8 & 15.6 & 12.47 & --- & --- & 11.2 & 16.2 & 18.3 \\
13 & J004818-730558  & 00:48:18.7 & -73:05:58 & --- & 4.9 & 2.15 & --- & --- & --- & --- & --- \\
14 & J004826-730606  & 00:48:27.0 & -73:06:06 & --- & 4.9 & 1.93 & --- & --- & --- & --- & --- \\\smallskip
15 & J004829-730626  & 00:48:29.9 & -73:06:26 & --- & 4.4 & 0.42 & --- & --- & --- & --- & --- \\
16 & J004836-725759  & 00:48:36.4 & -72:57:59 & --- & 2.8 & 1.59 & --- & --- & --- & 2.2 & --- \\
17 & J004841-732614  & 00:48:41.8 & -73:26:14 & --- & --- & 0.58 & --- & --- & --- & --- & --- \\
18 & J004857-730952  & 00:48:57.1 & -73:09:52 & --- & 1.6 & 0.91 & --- & --- & --- & --- & --- \\
19 & J004901-731109  & 00:49:01.8 & -73:11:09 & --- & --- & 0.45 & --- & --- & --- & --- & 1.6 \\\smallskip
20 & J004929-732633  & 00:49:29.2 & -73:26:33 & 9.7 & 7.2 & 5.53 & --- & --- & 4.1 & 5.9 & 5.6 \\
21 & J004930-732621  & 00:49:30.4 & -73:26:21 & --- & --- & 0.23 & --- & --- & --- & --- & --- \\
22 & J004940-730958  & 00:49:40.4 & -73:09:58 & --- & --- & 0.15 & --- & --- & --- & --- & --- \\
23 & J004941-724840  & 00:49:41.5 & -72:48:40 & 6.6 & 2.8 & 1.42 & --- & --- & 3.4 & 5.7 & 5.5 \\
24 & J004942-731037  & 00:49:42.7 & -73:10:37 & 7.3 & 3.6 & 3.25 & --- & --- & --- & 5.2 & 3.5 \\\smallskip
25 & J004946-731024  & 00:49:46.1 & -73:10:24 & --- & --- & 0.25 & --- & --- & --- & --- & --- \\
26 & J005043-724655  & 00:50:43.4 & -72:46:55 & --- & 7.1 & 0.53 & --- & --- & --- & --- & --- \\
27 & J005141-731331  & 00:51:41.2 & -73:13:31 & 12.5 & 7.7 & 2.88 & --- & --- & 5.7 & 10.8 & 8.0 \\
28 & J005148-725041  & 00:51:48.3 & -72:50:41 & 9.2 & 7.2 & 2.27 & --- & --- & 5.1 & 5.9 & 6.6 \\
29 & J005158-732030  & 00:51:58.3 & -73:20:30 & --- & --- & 0.32 & --- & --- & --- & --- & --- \\\smallskip
30 & J005212-733604  & 00:52:12.6 & -73:36:04 & --- & 1.4 & 0.20 & --- & --- & --- & 1.3 & --- \\
31 & J005729-723223  & 00:57:29.9 & -72:32:23 & 4.0 & 4.4 & --- & --- & --- & 1.4 & 3.3 & 3.3 \\
32 & J005816-723849  & 00:58:16.5 & -72:38:49 & 12.7 & 5.1 & --- & --- & --- & 5.0 & 14.8 & 9.5 \\
33 & J005914-721103  & 00:59:14.9 & -72:11:03 & --- & --- & --- & 2.2 & 5.9 & --- & --- & --- \\
34 & J005911-721140  & 00:59:11.0 & -72:11:40 & --- & --- & --- & --- & --- & --- & 7.0$\ast$ & 5.0$\ast$ \\\smallskip
35 & J010058-713527  & 01:00:58.5 & -71:35:27 & 9.3 & 6.7 & --- & --- & --- & 4.7 & 7.5 & 7.5 \\
36 & J010132-715042  & 01:01:32.2 & -71:50:42 & 1.5 & 2.4 & --- & --- & --- & --- & --- & --- \\
37 & J010243-715331  & 01:02:43.6 & -71:53:31 & 8.1 & 2.5 & --- & 0.4 & --- & --- & 2.6 & 1.6 \\
38 & J010248-715314  & 01:02:48.7 & -71:53:14 & 13.9 & 9.1 & --- & 5.9 & 6.9 & 5.3 & 8.9 & 12.1 \\
39 & J010503-715926  & 01:05:03.8 & -71:59:26 & --- & 14.5 & --- & 5.8 & 12.5 & --- & 24.5 & 17.4 \\\smallskip
40 & J010504-715900  & 01:05:05.0 & -71:59:00 & --- & 5.9 & --- & 2.5 & 3.0 & --- & 10.5 & 4.7 \\
41 & J010508-715946  & 01:05:08.4 & -71:59:46 & --- & 7.9 & --- & 3.1 & --- & --- & 10.3 & 6.2 \\
42 & J010832-721119  & 01:08:32.4 & -72:11:19 & --- & --- & --- & --- & --- & --- & 0.8$\ast$ & 0.6$\ast$ \\
43 & J010912-731138  & 01:09:13.0 & -73:11:38 & 50.0 & 33.1 & --- & --- & --- & 60.1 & 35.7 & 37.5 \\
44 & J011352-731546  & 01:13:52.1 & -73:15:46 & 8.7 & 3.5 & --- & --- & --- & --- & 4.7 & 4.8 \\\smallskip
45 & J011415-731550  & 01:14:15.8 & -73:15:50 & 9.5 & 4.3 & --- & --- & --- & 5.9 & 7.2 & 8.2 \\
46 & J011447-731946  & 01:14:47.1 & -73:19:46 & 15.0 & 8.0 & --- & --- & --- & 3.4 & 4.0 & 5.2 \\
47 & J011724-730917  & 01:17:25.0 & -73:09:17 & 20.1 & 12.5 & --- & --- & --- & 7.4 & 4.3 & 4.7 \\
48 & J012408-730904  & 01:24:08.1 & -73:09:04 & 39.6 & 51.4 & --- & --- & --- & 115.3 & 92.8 & 94.2 \\
\hline
\end{longtable}
$\ast$ Values taken from Indebetouw et al. 2004

%
\centerline{{\bf Table A2.}Compact \HII\ regions and their properties.}
\vskip1mm
\setlongtables
\small
\begin{longtable}{cccccccccccc}
\hline
(1)& (2)         & (3)        & (4)         & (5)           & (6)        & (7) & (8)\\
No & Name    & RA         & Dec       & S$_{1000}$  & Spectral Index & Luminosity& Luminosity in\\
   &             & (J2000)    & (J2000)& (mJy)&&$\times10^{26}$  W/Hz&$\mathscr{L}_\odot$\\
\hline
1 & J004312-725958  & 00:43:12.6 & -72:59:58 & --- & ---  & ---& ---  \\
2 & J004336-730227  & 00:43:36.4 & -73:02:27 & 4.4 & 0.1 $\pm$ 0.2  & 2.5& 0.6  \\
3 & J004451-725734  & 00:44:51.6 & -72:57:34 & --- & ---  & ---& ---  \\
4 & J004457-731012  & 00:44:57.1 & -73:10:12 & 4.0 & 0.0 $\pm$ 0.2  & 1.7  & 0.5 \\\smallskip
5 & J004502-731639  & 00:45:02.6 & -73:16:39 & 5.0 & -0.1 $\pm$ 0.2  & 1.7& 0.4  \\
6 & J004515-730607  & 00:45:15.7 & -73:06:07 & --- & ---  & ---& ---  \\
7 & J004610-732534  & 00:46:10.7 & -73:25:34 & --- & ---  & ---& ---  \\
8 & J004617-731243  & 00:46:17.7 & -73:12:43 & --- & ---  & ---& ---  \\
9 & J004640-732221  & 00:46:40.3 & -73:22:21 & --- & ---  & ---  & --- \\\smallskip
10 & J004645-731017  & 00:46:45.9 & -73:10:17 & --- & ---  & ---& ---  \\
11 & J004753-731709  & 00:47:53.5 & -73:17:09 & --- & ---  & ---& ---  \\
12 & J004808-731454  & 00:48:08.6 & -73:14:54 & 14.9 & 0.0 $\pm$ 0.1  & 7.1& 1.9  \\
13 & J004818-730558  & 00:48:18.7 & -73:05:58 & --- & ---  & ---& ---  \\
14 & J004826-730606  & 00:48:27.0 & -73:06:06 & --- & ---  & ---  & --- \\\smallskip
15 & J004829-730626  & 00:48:29.9 & -73:06:26 & --- & ---  & ---& ---  \\
16 & J004836-725759  & 00:48:36.4 & -72:57:59 & 2.1 & 0.1 $\pm$ 0.4  & 1.1& 0.3  \\
17 & J004841-732614  & 00:48:41.8 & -73:26:14 & --- & ---  & ---& ---  \\
18 & J004857-730952  & 00:48:57.1 & -73:09:52 & --- & ---  & ---& ---  \\
19 & J004901-731109  & 00:49:01.8 & -73:11:09 & 0.3 & 0.7 $\pm$ 0.1  & 2.2  & 0.6 \\\smallskip
20 & J004929-732633  & 00:49:29.2 & -73:26:33 & 7.0 & -0.2 $\pm$ 0.1  & 1.7& 0.4  \\
21 & J004930-732621  & 00:49:30.4 & -73:26:21 & --- & ---  & ---& ---  \\
22 & J004940-730958  & 00:49:40.4 & -73:09:58 & --- & ---  & ---& ---  \\
23 & J004941-724840  & 00:49:41.5 & -72:48:40 & 3.1 & 0.2 $\pm$ 0.3  & 3.1& 0.8  \\
24 & J004942-731037  & 00:49:42.7 & -73:10:37 & 4.8 & -0.1 $\pm$ 0.2  & 1.4  & 0.4 \\\smallskip
25 & J004946-731024  & 00:49:46.1 & -73:10:24 & --- & ---  & ---& ---  \\
26 & J005043-724655  & 00:50:43.4 & -72:46:55 & --- & ---  & ---& ---  \\
27 & J005141-731331  & 00:51:41.2 & -73:13:31 & 6.7 & 0.1 $\pm$ 0.3  & 3.9& 1.0  \\
28 & J005148-725041  & 00:51:48.3 & -72:50:41 & 5.5 & 0.0 $\pm$ 0.3  & 2.6& 0.7  \\
29 & J005158-732030  & 00:51:58.3 & -73:20:30 & --- & ---  & ---  & --- \\\smallskip
30 & J005212-733604  & 00:52:12.6 & -73:36:04 & 0.4 & 0.7 $\pm$ 1.4  & 2.8& 0.7  \\
31 & J005729-723223  & 00:57:29.9 & -72:32:23 & 3.3 & -0.1 $\pm$ 0.3  & 1.1& 0.3  \\
32 & J005816-723849  & 00:58:16.5 & -72:38:49 & 7.7 & 0.1 $\pm$ 0.3  & 5.1& 1.3  \\
33 & J005911-721140  & 00:59:11.0 & -72:11:40 & 17.3 & -0.6 $\pm$ 0.3  & 1.2& 0.3  \\
34 & J005914-721103  & 00:59:14.9 & -72:11:03 & 1.1 & 2.0 $\pm$ 0.4  & 1363.6  & 355.2 \\\smallskip
35 & J010058-713527  & 01:00:58.5 & -71:35:27 & 7.2 & -0.0 $\pm$ 0.1  & 2.7& 0.7  \\
36 & J010132-715042  & 01:01:32.2 & -71:50:42 & 1.7 & 1.0 $\pm$ 0.4  & 35.4& 9.2  \\
37 & J010243-715331  & 01:02:43.6 & -71:53:31 & 2.4 & -0.2 $\pm$ 0.7  & 0.5& 0.1  \\
38 & J010248-715314  & 01:02:48.7 & -71:53:14 & 8.1 & 0.0 $\pm$ 0.2  & 4.1& 1.1  \\
39 & J010503-715926  & 01:05:03.8 & -71:59:26 & 8.4 & 0.5 $\pm$ 0.3  & 20.2  & 5.3 \\\smallskip
40 & J010504-715900  & 01:05:05.0 & -71:59:00 & 3.4 & 0.3 $\pm$ 0.4  & 4.7& 1.2  \\
41 & J010508-715946  & 01:05:08.4 & -71:59:46 & 4.9 & 0.2 $\pm$ 0.4  & 5.0& 1.3  \\
42 & J010832-721119  & 01:08:32.4 & -72:11:19 & 1.7 & -0.5 $\pm$ 0.3  & 0.1& 0.0  \\
43 & J010912-731138  & 01:09:13.0 & -73:11:38 & 46.1 & -0.1 $\pm$ 0.1  & 14.2& 3.7  \\
44 & J011352-731546  & 01:13:52.1 & -73:15:46 & 5.9 & -0.1 $\pm$ 0.2  & 1.5  & 0.4 \\\smallskip
45 & J011415-731550  & 01:14:15.8 & -73:15:50 & 6.5 & 0.1 $\pm$ 0.2  & 3.3& 0.9  \\
46 & J011447-731946  & 01:14:47.1 & -73:19:46 & 9.3 & -0.5 $\pm$ 0.3  & 0.9& 0.2  \\
47 & J011724-730917  & 01:17:25.0 & -73:09:17 & 15.5 & -0.7 $\pm$ 0.1  & 0.9& 0.2  \\
48 & J012408-730904  & 01:24:08.1 & -73:09:04 & 51.2 & 0.4 $\pm$ 0.2  & 90.0& 23.4  \\
\hline
\end{longtable}
\end{landscape}
\vskip.5cm

\vskip.5cm

\vfill\eject

{\ }



\naslov{NOVO PROUQAVA{NJ}E MALOG MAGELANOVOG OBLAKA U RADIO-KONTINUMU NA 20~CM: DEO~{\bf III} - KOMPAKTNI \textrm{HII} REGIONI}


\authors{G. F. Wong, M. D. Filipovi\'c, E. J. Crawford,  N.~F.~H. Tothill, }
\authors{A.~Y. De Horta, T.~J.~Galvin}

\vskip3mm


\address{University of Western Sydney, Locked Bag 1797, Penrith South DC, NSW 2751, AUSTRALIA}

\Email{m.filipovic}{uws.edu.au}

\vskip3mm


\centerline{\rrm UDK \udc}

\vskip1mm

\centerline{\rit Originalni nauqni rad}

\vskip.7cm

\begin{multicols}{2}

{


\rrm 

Predstav{lj}amo novi katalog od 48 kompaktna \textrm{HII} regiona u Malom Magelanovom Oblaku (MMO). Tako{dj}e, predstav{lj}amo i nove radio kontinum slike \textrm{N19} regiona koji se nalazi u jugo-zapadnom delu MMO-a. Nove slike su kreirane spaja{nj}em svih raspolo{\zz}ivih 20~cm posmatra{nj}a sa \textrm{Australian Telescope Compact Array}. Ve{\cc}ina detektovanih komapkt \textrm{HII} regiona ima tipiqan ``ravan'' spektar {\ss}to je karakteristika dominantne termalne emisije.
}

\end{multicols}

\end{document}